\documentclass[shortnames,nojss]{jss}

\usepackage{thumbpdf,lmodern,setspace}
\usepackage{pbox}
\usepackage{array}
\usepackage{float}

\author{Robert C. Free\\University of Leicester}
\Plainauthor{Robert C. Free}

\title{Simpler handling of clinical concepts in R with clinconcept}
\Plaintitle{Simpler handling of clinical concepts in R with clinconcept}
\Shorttitle{clinconcept}

\Abstract{
    Routinely collected data in electronic healthcare records are often underpinned by clinical concept dictionaries. Increasingly data sets from these sources are being made available and used for research purposes, but without additional tooling it can be difficult to work effectively with these dictionaries due to their design, size and complex nature. In an effort to improve this situation the \pkg{clinconcept} package was created to provide a straightforward way for researchers to build, manage and interrogate databases containing commmonly used clinical concept dictionaries. This article describes the rationale behind the package, how to install it and use it and how it can be extended to support other data sources.
}

\Keywords{Read, SNOMED-CT, ICD10, clinical, concept, EHR, \proglang{R}, terminologies, data}
\Plainkeywords{Read, SNOMED-CT, ICD10, clinical, concept, EHR, R, terminologies, data}

\Address{
  Robert C. Free\\
  Department of Respiratory Sciences\\
  University of Leicester \\
  NIHR Leicester Biomedical Research Centre\\
  Respiratory\\
  Glenfield Hospital\\
  Groby Road\\
  Leicester\\
  LE3 9QP\\
  UK\\
  E-mail: \email{rob.free@le.ac.uk}\\
  URL: \url{https://rcfgroup.github.io}
}

\begin{document}

\section[Introduction]{Introduction} \label{sec:intro}

Routinely collected health care data are often recorded in electronic health care record (EHR) systems using clinical concepts. These concepts take the form of dictionaries containing controlled vocabularies which map a concept code to human-readable/searchable terms. Since different health-care workers use different descriptions for concepts, dictionaries may also contain synonyms for the same concept, which allows those workers to find concepts more easily. The structure of controlled vocabularies can vary from a simple two-level hierarchy such as the 10th revision of the International Statistical Classification of Diseases and Related Health Problems (ICD10) \citep{ICD10:2017} through to a more complicated diacyclic graph (DAG) such as the Systematized Nomenclature of Medicine Clinical Terms (SNOMED-CT) \citep{SNOMED:2010}. Depending on the complexity of the vocabulary structure, information can also be recorded at various levels of specificity. For example, a patient could be coded generically with Asthma, or be given a more specific diagnosis of Eosinophilic asthma). Additionally, the size of vocabularies can vary from tens of thousands to hundreds of thousands of terms, concepts and codes.

Clinically coded EHR data from primary and secondary care are increasingly being made available for research purposes. In the UK this not only includes large resources and cohorts such as the Clinical Practice Research Datalink \citep{cprd} and UK Biobank \citep{UK_biobank}, but also improved access and availability to smaller cohorts such as the Extended Cohort for E-health, Environment and DNA (EXCEED) \citep{exceed} through initiatives such as Health Data Research UK \citep{hdruk}. However without additional tooling it can be difficult for researchers to build, work with and interrogate clinical coding terminologies and dictionaries due to the design, size and complex nature of these dictionaries. In an effort to improve this situation \pkg{clinconcept} was created. This package helps researchers work with clinical concept dictionaries more effectively by providing functionality to i) build a local or remote database containing one or more clinical dictionaries, using UK NHS provided downloads of these dictionaries; ii) search clinical concepts using a \pkg{dplyr} focused interface to achieve complex searches; and iii) extract child and parent codes/terms with support for hierarchical and DAG-based searches (depending on the dictionary). Supported in the current version of the package are support for loading Read versions 2 and 3 \citep{ONeil:1995}, SNOMED-CT \citep{SNOMED:2010} and ICD10 \citep{ICD10:2017}.

\section{Installing the Package} \label{sec:install}

To install the latest version of \pkg{clinconcept} it should be installed from github using the \pkg{devtools} package:

\begin{CodeInput}
devtools::install_github("rcfgroup/clinconcept")
library(clinconcept)
\end{CodeInput}

\section{Building the Concept Database} \label{sec:building}

\pkg{clinconcept} builds concept databases using data files downloaded from the NHS Technology Reference data Update Distribution (NHS TRUD) web-site (https://isd.digital.nhs.uk). In order to obtain access to these downloads it is necessary to register with the NHS TRUD. For legal reasons it has not been not possible to include these files with the \pkg{clinconcept} package or make them available outside the NHS TRUD as separate downloads.

After you have registered, you will need to request access to, and then download your required code sets (see appendix A for details of the download types currently supported by \pkg{clinconcept}). Once downloaded, you should extract the files and make a note of the base location of these files for later use. As of 2020, Read version 2 and 3 downloads are deprecated and have been/are being withdrawn from the main TRUD site, but are available through a different route (see appendix A).

To build concept databases \pkg{clinconcept} uses a database management system (DBMS); currently it supports SQLite or MySQL. Of these, SQLite is the simplest to setup and use. If it is not already available on your local system it can easily be installed on Linux platforms through a package manager, or if using Windows by downloading it from the SQLite web site (https://sqlite.org/). If you plan to use MySQL as your DBMS you will also need to install the \pkg{RMySQL} package manually using install.packages.

Once the concept dictionary files have been downloaded and extracted, and a DBMS has been installed, a concept database can be built.

In this example we will use SQLite as DBMS to create a local build of the Read version 3 dictionaries; but the same approach should work for Read version 2,SNOMED-CT and ICD10.

A \pkg{clinconcept} clinical dictionary object must be created initially. This object provides a container for the configuration and database connection. To simplify this \code{cc\_from_} helper methods are provided.

If creating local SQLite databases the \code{cc\_from\_list} function can be used as no user credentials will be required, but if setting up a remote database it is recommended that the \code{cc\_from\_file} or \code{cc\_from\_home} functions are used due to the need for a username and password. The former loads options from the specified file path, while the latter loads a specified file from the user's home directory. The JSON file format is supported for this purpose and an example JSON file is given below. When using MySQL 'user', 'pass', 'port' and 'host' options may also be included.

\begin{CodeInput}
{
	"type":"sqlite",
	"dbname":"/path/to/readv3_db.sqlite"
}
\end{CodeInput}

The \code{cc\_from_} functions also require a concept dictionary type. Currently supported are \emph{NHSReadV2}, \emph{NHSReadV3}, \emph{NHSICD10} and \emph{NHSSnomedCT}.

All three of the approaches below are equivalent:
\begin{CodeInput}

library(clinconcept)

#read options from list (not recommended for MySQL databases)
dict<-cc_from_list("NHSReadV3", list(type =
 "sqlite",name = "/path/to/readv3_db.sqlite"))

#read readv3.cfg from current user's home directory
dict<-cc_from_home("NHSReadV3", "readv3.cfg")

#read specific file path: /path/to/config/readv3.cfg
dict<-cc_from_file("NHSReadV3", "/home/user/readv3.cfg")
\end{CodeInput}

We can now build concept databases using the clinical dictionary object. In our example, Read version 3 data files have been downloaded from the NHS TRUD site and extracted into /home/user/ReadV3.

\begin{CodeInput}
build_concept_tables(dict,'/home/user/ReadV3')
\end{CodeInput}

Loading concepts may take some time, however it only needs to be done once for each concept type.

\section{Term and Concept Searches} \label{sec:tcsearches}
Once dictionary building is complete, advanced searches can be carried out using the \code{search\_concept} function. This function wraps \pkg{dplyr}'s filter function meaning these searches can be carried out using \proglang{R} syntax. Some examples are given below:

\begin{CodeInput}
#find all concepts containing the term 'Asthma':
asthma<-search_concepts(dict, term == "Asthma")

#find concepts containing the term 'Asthma' but excluding 'Eosinophilic asthma':
asthma<-search_concepts(dict, term == "Asthma" & (! term == "Eosinophilic asthma")
\end{CodeInput}

Due to the nature of clinical concept codes it is important to ensure that case-sensitivity is taken into account. Unfortunately SQLite queries are by default not case-sensitive, which can cause incorrect concepts to be returned. To manage this, \pkg{clinconcept} provides utility functions to enable and disable case sensitivity which should be included at appropriate points in programs. It is not necessary to use these functions if using a DBMS other than SQLite, however if present they will be ignored.

\begin{CodeInput}
enable_case_sensitivity(dict)

#find concepts with code H3...
search_concepts(dict, read_code == "H3...")

#find concepts with codes beginning with H3
search_concepts(dict, read_code %like% "H3%")

#two ways to find concepts with code equal to H3... or H31...
search_concepts(dict, read_code %in% ("H3...","H31..")
search_concepts(dict, read_code == "H3..." | read_code == "H31..")

disable_case_sensitivity(dict)
\end{CodeInput}

The output type from the \code{search_concepts} function is specified through the output argument. The default is a \pkg{dplyr} \code{tbl} object which can be further processed using \pkg{dplyr} functions (e.g. group) or finalised into a data.frame-like object (a tibble) using \code{dplyr::collect}, but it is also possible to return terms and codes as vectors, with duplicate codes removed from the latter:

\begin{CodeInput}
> search_concepts(dict, term %like% "%airway%")
\end{CodeInput}

\begin{CodeOutput}
# Source:   lazy query [?? x 8]
# Database: sqlite 3.19.3
#   [/home/user/clindata/readv3.sqlite]
  read_code                                                                    term
      <chr>                                                                   <chr>
1     H3122               Acute exacerbation of chronic obstructive airways disease
2     H3y..                     Other specified chronic obstructive airways disease
3     H3z..                                 Chronic obstructive airways disease NOS
4     Xa35l "Acute infective exacerbation of chronic obstructive airways disease
5     XaIND                           End stage chronic obstructive airways disease
# ... with 6 more variables: term_30 <chr>, term_60 <chr>, term_198 <chr>, term_id
# <chr>, synonym <chr>, status <chr>
\end{CodeOutput}

\begin{CodeInput}
> search_concepts(dict, term %like% "%airway%", output="terms")
\end{CodeInput}

\begin{CodeOutput}
[1] "Acute exacerbation of chronic obstructive airways disease"
[2] "Other specified chronic obstructive airways disease"
[3] "Chronic obstructive airways disease NOS"
[4] "Acute infective exacerbation of chronic obstructive airways disease"
[5] "End stage chronic obstructive airways disease"
\end{CodeOutput}

\begin{CodeInput}
> search_concepts(dict, term %like% "%airway%", output="codes")
\end{CodeInput}

\begin{CodeOutput}
[1] "H3122" "H3y.." "H3z.." "Xa35l" "XaIND"
\end{CodeOutput}

The default is to exclude synonyms from searches, but these can be included using the \code{include\_synonyms} parameter:

\begin{CodeInput}
> search_concepts(dict, read_code == "H3...", output="terms")
\end{CodeInput}

\begin{CodeOutput}
 [1] "Chronic obstructive lung disease"
\end{CodeOutput}

\begin{CodeInput}
> search_concepts(dict, read_code == "H3...", include_synonyms=T, output="terms")
\end{CodeInput}

\begin{CodeOutput}
 [1] "Chronic obstructive lung disease"
 [2] "COLD - Chronic obstructive lung disease"
 [3] "Chronic obstructive pulmonary disease"
 [4] "COPD - Chronic obstructive pulmonary disease"
 [5] "Chronic obstructive airway disease"
 [6] "COAD - Chronic obstructive airways disease"
 [7] "Chronic obstructive bronchitis"
 [8] "Chronic airway disease"
 [9] "Chronic airway obstruction"
[10] "Chronic airflow limitation"
[11] "Chronic airflow obstruction"
[12] "Chronic irreversible airway obstruction"
[13] "Obstructive chronic bronchitis"
[14] "COB - Chronic obstructive bronchitis"
[15] "CAFL - Chronic airflow limitation"
[16] "CAL - Chronic airflow limitation"
[17] "CAO - Chronic airflow obstruction"
\end{CodeOutput}

\section{Relationship Searches} \label{sec:relsearch}
The \code{search_concepts} function allows linear searches, but does not take into account the relationship between codes. To achieve this \pkg{clinconcept} provides functions to retrieve child and parent codes.

For instance the following examples will return all descendents (e.g. children, grand children etc) and immediate children:

\begin{CodeInput}
> get_child_codes(dict,"H3...")
\end{CodeInput}

\begin{CodeOutput}
 [1] "H3122" "H312z" "H3y.." "H3y0." "H3z.." "H4641" "Hyu31" "X101i" "X101l"
[10] "X101m" "X102z" "Xa35l" "XaEIV" "XaEIW" "XaEIY" "XaIND" "XaN4a" "XaZd1"
\end{CodeOutput}

\begin{CodeInput}
> get_child_codes(dict,"H3...",immediate_children = T)
\end{CodeInput}

\begin{CodeOutput}
 [1] "H3122" "H312z" "H3y.." "H3z.." "Hyu31" "X101l" "XaEIV" "XaEIW" "XaEIY"
[10] "XaIND" "XaN4a"
\end{CodeOutput}

You can also get all the ancestor codes for a concept, including the immediate parents.

\begin{CodeInput}
> get_parent_codes(dict, "H32..")
\end{CodeInput}

\begin{CodeOutput}
[1] "....." "H...." "X0003" "XaBVJ"
\end{CodeOutput}

\begin{CodeInput}
> get_parent_codes(dict, "H32..", immediate_parents = T)
\end{CodeInput}

\begin{CodeOutput}
[1] "H...."
\end{CodeOutput}

These commands return a vector of codes which can be used as an argument in \code{search\_concepts}:

\begin{CodeInput}
> h32_ancestors<-get_parent_codes(dict, "H32..")
> search_concepts(dict,read_code %in% h32_ancestors)
\end{CodeInput}

\begin{CodeOutput}
  read_code                 term              term_30 term_60 term_198 term_id
      <chr>                <chr>                <chr>   <lgl>    <lgl>   <chr>
1     .....       Read thesaurus       Read thesaurus      NA       NA   00000
2     H.... Respiratory disorder Respiratory disorder      NA       NA   Y1001
3     X0003            Disorders            Disorders      NA       NA   Y006t
4     XaBVJ    Clinical findings    Clinical findings      NA       NA   Yaap0
# ... with 2 more variables: synonym <chr>, status <chr>
\end{CodeOutput}

\section{Technical Details}\label{sec:techdet}

\subsection{Algorithms}
The algorithm used internally by the relationship code search is dependant on the complexity of the dictionary and whether it is hierarchical or a DAG. In the case of Read version 2 which is a hierarchy, searches are done based on the first letters in the supplied code. For instance, to obtain child codes for H3... a database query is run for all those codes beginning with 'H3'. DAG-based searches, such as those used by SNOMED-CT use a recursive algorithm which finds ancestors/descendants at each level, and then retrieves each of their ancestors/descendants until no ancestors/descendents are returned. To improve performance the algorithm has been designed so that the same ancestors/descendents are not retrieved multiple times.

\subsection{Regression Tests}
The package contains regression tests based on \pkg{testthat} and some example concept data files. To run these tests SQLite must be pre-installed (the tests will check for the sqlite3 binary). Therefore, it is recommended that SQLite is installed before \pkg{clinconcept} is installed, so that the tests are run during package install. If not present, the tests are skipped automatically, although it is possible to run them following package installation by executing:

 \begin{CodeInput}
 devtools::test('clinconcept')
 \end{CodeInput}

The database table structure varies between different clinical dictionaries. The appendix includes additional information on the table structures and contents.

\subsection{Extending clinconcept }

If you have a clinical dictionary or DBMS which is not supported, \pkg{clinconcept} can be extended with support for these. It is recommended that you adapt one of the dict.R files to achieve this.

A brief example is given here for SQLite and a dictionary type called TESTCONCEPT. \pkg{clinconcept} uses S3 function calls with the clinical dictionary object (defined with both clinconcept and TESTCONCEPT classes). This means that the base clinconcept function will be called unless there is a TESTCONCEPT specific function. The building process is different as this is both dictionary AND DBMS specific.

The first requirement is to create the clinical dictionary object. This simply requires use of one of the \code{cc\_from\_} functions:

\begin{CodeInput}
dict<-cc_from_home("TESTCONCEPT", "/path/to/dictconfig.json")
\end{CodeInput}

Using a combination of SQL statements and R code means \pkg{clinconcept} can support building different types of concept dictionaries on different DBMS platforms. The default \\ \code{build\_concept\_tables} function executes the DBMS/dictionary-specific SQL statement files, the location of which is defined by \code{dict\$sql\_path}).  SQL files before and after a specific call to \\
\code{build\_concept\_tables.sqlite.TESTCONCEPT} (if it existed).

A series of dictionary specific get functions are also required. These return the table names, code and term field names:
\begin{CodeInput}
get_ctable_name.TESTCONCEPT<-function(dict) {"test_concept"}
get_ctable_code_field.TESTCONCEPT<-function(dict) {"concept_code"}
get_ctable_term_field.TESTCONCEPT<-function(dict) {"term"}
get_ptable_name.TESTCONCEPT<-function(dict) {"test_concept_parents"}
get_ptable_code_field.TESTCONCEPT<-function(dict) {"concept_code"}
get_ptable_parent_field.TESTCONCEPT<-function(dict) {"parent_concept_code"}
\end{CodeInput}

The default \code{search\_concepts} function uses \pkg{dplyr} and the table and field names returned by the get functions to search and provide output, so it may not be necessary to create a dictionary-specific version of this, although it can be done by creating a \code{search\_concepts.TESTCONCEPT} function.

The \code{get\_child\_codes.TESTCONCEPT} \code{get\_parent\_codes.TESTCONCEPT} functions will need to be created, although if the dictionary is a DAG, helper functions are available to simplify this procedure.


\section{Summary and Discussion} \label{sec:summary}

This article provides a general overview of the R package \pkg{clinconcept} which was created to simplify the use of routine health care data. A simpler approach to deal with data of this type is timely given the increased interest and availability of this type of data.

Some examples based on the use of Read version 3 are included to give the reader an idea of the overall building process and possible queries possible. Additional information for other types of clinical dictionary can be found in the appendix and in the R package documentation.

\pkg{clinconcept} provides a more extendable and straightforward build and query tool than the alternatives. Packages such as \pkg{rpcdsearch} \citep{rpcdsearch:2016}, while they have strengths do not support simple loading of NHS-based code lists or the ability to perform relationship-based searches. Non-R options such as graphical and web-based browsers, provide the required data but do not allow the results of these to be embedded into R scripts and used programmatically.

Future plans include adding support for drug concepts (Drugs and Medicines + Devices and British National Formulary) and extending the features of \pkg{clinconcept} to providing cross-mapping functionality. This should be straightforward because files to do this are available as NHS TRUD downloads. This would allow a researcher to determine which codes from one clinical concept dictionary are equivalent to codes in another dictionary. There is also the possibility of extending support for searches of SNOMED-CT concepts to include attributes.

It is hoped this package is useful to others and any suggestions for improvements or new features would be gratefully received.

\section*{Acknowledgments}

RCF developed the package and wrote the paper.
This work was supported by the National Institute for Health Research Leicester Biomedical Research Centre. The views expressed are those of the author and not necessarily those of the NHS, the National Institute for Health Research or the Department of Health.


\bibliography{refs}

\begin{thebibliography}{8}
\newcommand{\enquote}[1]{``#1''}
\providecommand{\natexlab}[1]{#1}
\providecommand{\url}[1]{\texttt{#1}}
\providecommand{\urlprefix}{URL }
\expandafter\ifx\csname urlstyle\endcsname\relax
  \providecommand{\doi}[1]{doi:\discretionary{}{}{}#1}\else
  \providecommand{\doi}{doi:\discretionary{}{}{}\begingroup
  \urlstyle{rm}\Url}\fi
\providecommand{\eprint}[2][]{\url{#2}}

\bibitem[{CPRD(2012)}]{cprd}
CPRD (2012).
\newblock \enquote{Clinical Practice Research Datalink.}
\newblock \url{https://www.cprd.com/}.
\newblock Accessed: 26/02/2020.

\bibitem[{HDR-UK(2017)}]{hdruk}
HDR-UK (2017).
\newblock \enquote{Health Data Research UK.}
\newblock \url{https://www.hdruk.ac.uk/}.
\newblock Accessed: 26/02/2020.

\bibitem[{{IHTSDO}(2010)}]{SNOMED:2010}
{IHTSDO} (2010).
\newblock \enquote{{SNOMED Clinical Terms User Guide}.}
\newblock \emph{Development}, (January), 99.

\bibitem[{John \emph{et~al.}(2019)John, Reeve, Free, Williams, Ntalla, Farmaki,
  Bethea, Barton, Shrine, Batini, and et~al.}]{exceed}
John C, Reeve NF, Free RC, Williams AT, Ntalla I, Farmaki AE, Bethea J, Barton
  LM, Shrine N, Batini C, et~al (2019).
\newblock \enquote{Cohort Profile: Extended Cohort for E-health, Environment
  and DNA (EXCEED).}
\newblock \emph{International Journal of Epidemiology}, \textbf{48}(3),
  678–679j.
\newblock ISSN 0300-5771.
\newblock \doi{10.1093/ije/dyz073}.

\bibitem[{Olier \emph{et~al.}(2016)Olier, Springate, Ashcroft, Doran, Reeves,
  Planner, Reilly, and Kontopantelis}]{rpcdsearch:2016}
Olier I, Springate DA, Ashcroft DM, Doran T, Reeves D, Planner C, Reilly S,
  Kontopantelis E (2016).
\newblock \enquote{Modelling Conditions and Health Care Processes in Electronic
  Health Records: An Application to Severe Mental Illness with the Clinical
  Practice Research Datalink.}
\newblock \emph{PLOS ONE}, \textbf{11}(2), e0146715.
\newblock ISSN 1932-6203.
\newblock \doi{10.1371/journal.pone.0146715}.

\bibitem[{O'Neil \emph{et~al.}(1995)O'Neil, Payne, and Read}]{ONeil:1995}
O'Neil M, Payne C, Read J (1995).
\newblock \enquote{{Read Codes Version 3: a user led terminology.}}
\newblock \emph{Methods of information in medicine}, \textbf{34}(1-2), 187--92.
\newblock ISSN 0026-1270.
\newblock \urlprefix\url{http://www.ncbi.nlm.nih.gov/pubmed/9082130}.

\bibitem[{Sudlow \emph{et~al.}(2015)Sudlow, Gallacher, Allen, Beral, Burton,
  Danesh, Downey, Elliott, Green, Landray, and et~al.}]{UK_biobank}
Sudlow C, Gallacher J, Allen N, Beral V, Burton P, Danesh J, Downey P, Elliott
  P, Green J, Landray M, et~al (2015).
\newblock \enquote{UK Biobank: An Open Access Resource for Identifying the
  Causes of a Wide Range of Complex Diseases of Middle and Old Age.}
\newblock \emph{PLoS Medicine}, \textbf{12}(3).
\newblock ISSN 1549-1277.
\newblock \doi{10.1371/journal.pmed.1001779}.
\newblock
  \urlprefix\url{https://www.ncbi.nlm.nih.gov/pmc/articles/PMC4380465/}.

\bibitem[{WHO(2017)}]{ICD10:2017}
WHO (2017).
\newblock \emph{{Classification of Diseases (ICD)}}.
\newblock \urlprefix\url{http://www.who.int/classifications/icd/en/}.

\end{thebibliography}


\newpage
\begin{appendix}

\section{Information about Supported Clinical Dictionaries}

The table below contains information on the supported downloads and the queryable fields for each type of clinical dictionary to help when querying with the \code{search\_concepts} function. \pkg{clinconcept} dictionary types are included in parentheses in the \emph{Dictionary} column.

\begin{tabular}{| >{\raggedright}p{3cm} | >{\raggedright}p{7cm} | l  | }
\hline
  Dictionary & Fields & Download name \\ \hline
  ICD10 (NHSICD10) & \pbox{7cm}{
  \textbf{icd10\_code}: ICD10 version 5 code \\
  \textbf{term}: human-readable/searchable clinical term \\
  \textbf{description}: original description \\
  \textbf{modifier\_4}: modifier text added to description to get term based on 4th character in code \\
  \textbf{modifier\_5}  modifier text added to description to get term based on 5th character in code \\
  \textbf{tree\_description}: extended description of term \\
  } & \pbox{4cm}{ICD10 5th Edition data file release}\\ \hline
  Read V2 (NHSReadV2)* & \pbox{7cm}{
  \textbf{read\_code}: Read version 2 code (4-byte) \\
  \textbf{term}: human-readable/searchable clinical term \\
  } & NHS UK Read Codes Version 2*  \\ \hline
  Read V3/Clinical Terms Version 3 (NHSReadV3)* & \pbox{7cm}{
  \textbf{read\_code}: Read version 3 code (5-byte) \\
  \textbf{term}: human-readable/searchable clinical term \\
  \textbf{status}: current (C), redundant (R), optional (O) or extinct (E) \\
  \textbf{synonym}: true (1) or false (0) \\
  \textbf{term\_id}: unique term ID (a term may have the same read\_code, but it would have a different term ID)
   } & \pbox{4cm}{NHS UK Read Codes Clinical Terms Version 3} \\ \hline
   SNOMED-CT (NHSSnomedCT) & \pbox{7cm}{
  \textbf{snomed\_code}: SNOMED-CT code \\
  \textbf{term}: human-readable/searchable clinical term \\
  \textbf{status}: current (C), redundant (R), optional (O) or extinct (E) \\
  \textbf{synonym}: true (1) or false (0) \\
  \textbf{term\_id}: unique term ID (a term may have the same read\_code, but it would have a different term ID)
   } & \pbox{4cm}{UK SNOMED CT Clinical Edition, RF1} \\
   \hline
\end{tabular}
*Read version 2 and 3 now have to be obtained through a separate 'Retired Products' site.
An account to access this site can be obtained by emailing the information.standards@nhs.net email address with a request for access.
The downloaded ZIP files remain the same as those which were available through the TRUD site.
\newpage
\end{appendix}
\end{document}